\documentclass[aps,prl,nobibnotes,twocolumn,superscriptaddress,shortbibliography]{revtex4-1}

\usepackage{amsfonts}
\usepackage{mathrsfs}
\usepackage{amsmath}
\usepackage{color}
\usepackage{graphicx}
\usepackage{bm}
\usepackage{amssymb}
\usepackage{xspace}
\usepackage{epstopdf}
\usepackage{dcolumn}
\usepackage{longtable}
\usepackage{multirow}
\usepackage{float}
\usepackage{comment}
\usepackage{lineno}

\usepackage[colorlinks=true, letterpaper=true, pdfstartview=FitV,  linkcolor=blue, citecolor=blue, urlcolor=blue]{hyperref}

\makeatother

\begin{document}
\title{Hourglass Charge-Three Weyl Phonons}
\author{Xiaotian Wang}\thanks{X.W. and F.Z. contributed equally to this manuscript}
\email{xiaotianwang@swu.edu.cn}
\address{School of Physical Science and Technology, Southwest University, Chongqing 400715, China.}

\author{Feng Zhou}\thanks{X.W. and F.Z. contributed equally to this manuscript}
\address{School of Physical Science and Technology, Southwest University, Chongqing 400715, China.}

\author{Zeying Zhang}
\affiliation{College of Mathematics and Physics, Beijing University of Chemical Technology, Beijing 100029, China}
\affiliation{Research Laboratory for Quantum Materials, Singapore University of
Technology and Design, Singapore 487372, Singapore}

\author{Zhi-Ming Yu}\email{zhiming$_$yu@bit.edu.cn}
\affiliation{Centre for Quantum Physics, Key Laboratory of Advanced Optoelectronic Quantum Architecture and Measurement (MOE), School of Physics, Beijing Institute of Technology, Beijing, 100081, China}
\affiliation{Beijing Key Lab of Nanophotonics \& Ultrafine Optoelectronic Systems, School of Physics, Beijing Institute of Technology, Beijing, 100081, China}
\author{Yugui Yao}
\affiliation{Centre for Quantum Physics, Key Laboratory of Advanced Optoelectronic Quantum Architecture and Measurement (MOE), School of Physics, Beijing Institute of Technology, Beijing, 100081, China}
\affiliation{Beijing Key Lab of Nanophotonics \& Ultrafine Optoelectronic Systems, School of Physics, Beijing Institute of Technology, Beijing, 100081, China}

\begin{abstract}
Unconventional Weyl point  with nonlinear dispersion features higher topological charge
$|{\cal{C}}|>1$ and multiple topologically protected Fermi arc states at its boundary.
As a novel topological state, it has been attracting widespread attention.
However,  the unconventional Weyl point with   $|{\cal{C}}|=3$ has not yet been reported in realistic materials, even though it has been theoretically proposed for more than a decade.
In this work, based on first-principles calculations and theoretical analysis, we predict the existing material, $\rm\alpha$-LiIO$_3$ as the first realistic example with this unconventional Weyl point.
Specifically, in the phonon spectra of $\rm\alpha$-LiIO$_3$, two Weyl points with   ${\cal{C}}=-3$, connected by time-reversal symmetry, appear at the neck crossing-point of an hourglass-type band, leading to two hourglass charge-3 Weyl phonons.
The symmetry protection and the associated novel triple- and sextuple-helicoid surface arc states of the hourglass charge-3 Weyl phonons are revealed.
Our results uncover a hidden topological character of $\rm\alpha$-LiIO$_3$ and also show that the phonon spectra is a great platform for exploring  unconventional  topological states.
\end{abstract}
\maketitle

In the past decades, one of the most important findings in condensed matter physics is the prediction of topological Weyl semimetal~\cite{Wan2011Topological-PRB,RevModPhys.90.015001}, which shows that elementary  particles like Weyl fermion can emerge as low-energy excitations in topological semimetal materials,  opening the door for simulating  interesting phenomena in astrophysics and general relativity  on laboratory tables \cite{RN1562}.
The conventional Weyl point (WP) exhibits  relativistic linear dispersion along any direction in momentum space and carries unit topological charge (Chern number) $|{\cal{C}}|=1$~\cite{Wan2011Topological-PRB}.
Particularly, it is topologically protected and can exist in three-dimensional (3D) crystals without any space group  symmetry (except translation symmetry).
Many realistic materials have been predicted as topological Weyl semimetals with conventional WP ~\cite{weng2015weyl, yang2015weyl,huang2015weyl,soluyanov2015type,PhysRevB.92.161107,PhysRevLett.116.226801,PhysRevLett.117.066402,PhysRevB.94.165201,
PhysRevLett.117.236401,PhysRevB.97.060406,PhysRevB.97.235416,chang2016strongly,PhysRevB.93.201101}, and some of them have been experimentally confirmed ~\cite{lv2015observation,ARPES_PRX_DH_2015,doi:10.1126/science.aaa9297,deng2016experimental,PhysRevB.94.121113}.

\begin{table*}[t]
\centering
\small
\renewcommand\arraystretch{1.3}
  \caption{The candidate SGs that can host (hourglass) C-3 WP in spinless systems. Irreps denotes the irreducible (co-) representation of the little group associated to the  (hourglass) C-3 WP.}
  \label{table I}
  \begin{tabular*}{1\textwidth}{@{\extracolsep{\fill}}llllll}
  \hline
  \hline
  SG No. & SG symbol & Generators &Location & Irreps   & Species  \\
  \hline
  168 & P6 &  $\{C_{6z}|000\}$, ${\cal{T}}$  & $\Gamma$-A path&  $\{R_1,R_4\}$, $\{R_2,R_5\}$, $\{R_3,R_6\}$ & C-3 WP\\	
  169 & P6$_1$ & $\{C_{6z}|00 \frac{1}{6}\}$, ${\cal{T}}$  & $\Gamma$-A path&  $\{R_1,R_4\}$, $\{R_2,R_5\}$, $\{R_3,R_6\}$ & C-3 WP\\
  170 & P6$_5$	&  $\{C_{6z}|00 \frac{5}{6}\}$, ${\cal{T}}$  & $\Gamma$-A path&  $\{R_1,R_4\}$, $\{R_2,R_5\}$, $\{R_3,R_6\}$ & C-3 WP\\
  171 & P6$_2$ &  $\{C_{6z}|00 \frac{1}{3}\}$, ${\cal{T}}$  & $\Gamma$-A path&  $\{R_1,R_4\}$, $\{R_2,R_5\}$, $\{R_3,R_6\}$ & C-3 WP\\
  172 & P6$_4$ &  $\{C_{6z}|00 \frac{2}{3}\}$, ${\cal{T}}$  & $\Gamma$-A path&  $\{R_1,R_4\}$, $\{R_2,R_5\}$, $\{R_3,R_6\}$ & C-3 WP\\
    173 & P6$_3$ &  $\{C_{6z}|00 \frac{1}{2}\}$, ${\cal{T}}$  & $\Gamma$-A path&  $\{R_3,R_6\}$ &  C-3 WP\\
    &  &     &  &  $\{R_1,R_4\}$, $\{R_2,R_5\}$  & Hourglass C-3 WP\\
  177 & P622 &  $\{C_{6z}|000\}$,$\{C_{21}^{'}|000\}$, ${\cal{T}}$  & $\Gamma$-A path&  $\{R_1,R_4\}$, $\{R_2,R_5\}$, $\{R_3,R_6\}$   & C-3 WP\\
  178 & P6$_1$22 &  $\{C_{6z}|00 \frac{1}{6}\}$,$\{C_{21}^{'}|000\}$, ${\cal{T}}$  & $\Gamma$-A path&  $\{R_1,R_4\}$, $\{R_2,R_5\}$, $\{R_3,R_6\}$  & C-3 WP\\
  179 & P6$_5$22 &  $\{C_{6z}|00\frac{5}{6}\}$, $\{C_{21}^{'}|000\}$,${\cal{T}}$  & $\Gamma$-A path&  $\{R_1,R_4\}$, $\{R_2,R_5\}$, $\{R_3,R_6\}$  & C-3 WP\\
  180 & P6$_2$22 &  $\{C_{6z}|00\frac{1}{3}\}$, $\{C_{21}^{'}|000\}$,${\cal{T}}$  & $\Gamma$-A path&  $\{R_1,R_4\}$, $\{R_2,R_5\}$, $\{R_3,R_6\}$  & C-3 WP\\
  181 & P6$_4$22 &  $\{C_{6z}|00\frac{2}{3}\}$, $\{C_{21}^{'}|000\}$,${\cal{T}}$  & $\Gamma$-A path&  $\{R_1,R_4\}$, $\{R_2,R_5\}$, $\{R_3,R_6\}$  & C-3 WP\\
  182 & P6$_3$22 &  $\{C_{6z}|00\frac{1}{2}\}$, $\{C_{21}^{'}|000\}$,${\cal{T}}$  & $\Gamma$-A path&  $\{R_3,R_6\}$,  &  C-3 WP\\	
    &   &     &  &   $\{R_1,R_4\}$, $\{R_2,R_5\}$  & Hourglass C-3 WP\\	
  \hline
  \hline
  \end{tabular*}
\end{table*}

In 2012, Fang \emph{et al.}~\cite{Fang2012Multi-Prl} demonstrated that with  rotation symmetry, two (three) conventional WPs can merge together, leading to unconventional WP, which exhibits higher topological charge $|{\cal{C}}|=2$ ($|{\cal{C}}|=3$)  and quadratic (cubic) dispersion in the plane normal to the rotation axis.
This pioneering work led to a series of subsequent studies,
such as the prediction of the corresponding  material candidates~\cite{chang2016room,huang2016new,PhysRevB.97.054305,PhysRevB.102.195104,he2020observation,PhysRevB.104.205124,PhysRevB.105.075133,PhysRevB.106.085102},  the demonstration of the novel phenomena of these unconventional WPs~\cite{PhysRevB.88.104412,PhysRevB.93.155125,PhysRevB.95.161112,PhysRevB.95.161306,PhysRevResearch.2.013007} and  the search of other types of unconventional emergent particles~\cite{yang2014classification,yang2015topological,gao2016classification,PhysRevB.99.121106,PhysRevB.101.205134}.
Recently, one other species of the unconventional WP, exhibiting topological charge $|{\cal{C}}|=4$, has been predicted in spinless nonmagnetic systems ~\cite{YU2022375,PhysRevB.102.125148,liu2020symmetry,PhysRevB.103.L161303} and spinful magnetic systems \cite{RN2550}.
The WP with topological charge $|{\cal{C}}|=n$ ($n=1,2,3,4$) also is  termed as charge-$n$ (C-$n$) WP \cite{YU2022375}.

However, up to now only a few materials were predicted to host the unconventional WPs. Particularly, the C-3 WP has not yet been reported in realistic material.
In the beginning, the topological Weyl semimetals are  predicted in the materials with strong spin-orbit coupling (SOC)~\cite{Wan2011Topological-PRB,Xu2011Chern-Prl,PhysRevB.89.081106,WengPRX_WP_2015}. Later, it was shown that spinless systems also can host WPs due to the pseudo-SOC effect \cite{YU2022375}, which significantly  expands the material candidates database of topological Weyl states. Recently,  the search of  WPs has shifted toward the phonon spectra of crystals~\cite{PhysRevLett.121.035302,PhysRevLett.120.016401,PhysRevLett.123.065501,PhysRevB.103.165143,PhysRevB.103.104101,li2021computation,PhysRevLett.124.105303,liu2022ubiquitous}.
Contrast to the electronic bands,  all the phonon bands are relevant for experimental detection, as the phonons are free of the constraint of Pauli exclusion principle and Fermi surface.
Additionally, the topological phonons may lead to novel phenomena in heat transfer, phonon scattering, and electron-phonon interaction~\cite{RevModPhys.84.1045,liu2020topological,PhysRevMaterials.2.114204,chen2021topological}.

In this work, based on theoretical analysis and first-principles calculations, we predict the  existing material, $\rm\alpha$-LiIO$_3$ as the first realistic example hosting the  C-3 WPs in its phonon spectra.
We first show that for spinless systems (like phonon spectra) with time-reversal symmetry ${\cal{T}}$, the  C-3 WP only appears at  sixfold  rotation axis belonging to chiral space groups (SGs). Hence, this unconventional WP can be realized in 12 of the 230 SGs, as listed in Table \ref{table I}. Moreover, the C-3 WP can be further divided into two categories: normal  C-3 WP  and hourglass  C-3 WP, which respectively appear in the SGs without and with sixfold screw rotation symmetry $\{C_{6z}|00\frac{1}{2}\}$. This means that the hourglass C-3 WP only appears in SGs 173 and 182.
Guided by the symmetry analysis, we identify the existence of the hourglass C-3 Weyl phonons in several realistic material candidates, including $\rm\alpha$-LiIO$_3$.
More material candidates can be found in Supplemental Material (SM) \cite{SM}.
Remarkably, the hourglass-type band and the hourglass C-3 WPs in $\rm\alpha$-LiIO$_3$ are well separated from other phonon bands.
Furthermore, we find that due to the presence of ${\cal{T}}$ symmetry, the hourglass C-3 WPs must come in pairs with the same chirality, leading to novel sextuple-helicoid surface arc state on the boundary normal to the sixfold screw rotation axis.
Our results not only predict the first material candidate hosting C-3 Weyl phonons, but also show that the phonon spectra is a great platform for exploring unconventional topological states.

\begin{figure}[b]
\includegraphics[width=8.5cm]{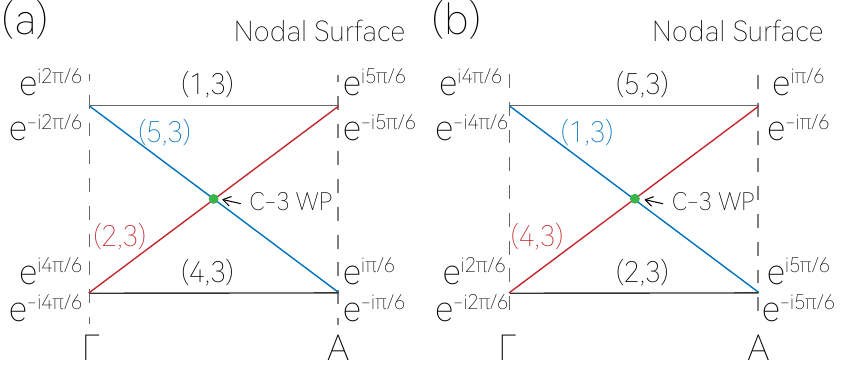}
\caption{Formation mechanism of hourglass C-3 WP in spinless systems. (a) and (b) show two possible modes of the hourglass-type bands along $\Gamma$-A path.
\label{fig1}}
\end{figure}

\emph{\textcolor{blue}{Hourglass C-3 WP.}}
As shown in previous works~\cite{Fang2012Multi-Prl,YU2022375}, for  spinless systems with time-reversal symmetry $\cal{T}$, the C-3 WP only occurs on the sixfold rotation axis of chiral SGs. Moreover, it can not reside  at the time-reversal-invariant momentum, as the topological charge of the WP at such momentum must be even \cite{arXiv.2203.13974}.
Hence, the C-3 WP should be a crossing that appears on $\Gamma$-A high-symmetry path and  is formed by two bands with different eigenvalue of the sixfold rotation operator.

\begin{figure}[b]
\includegraphics[width=8.5cm]{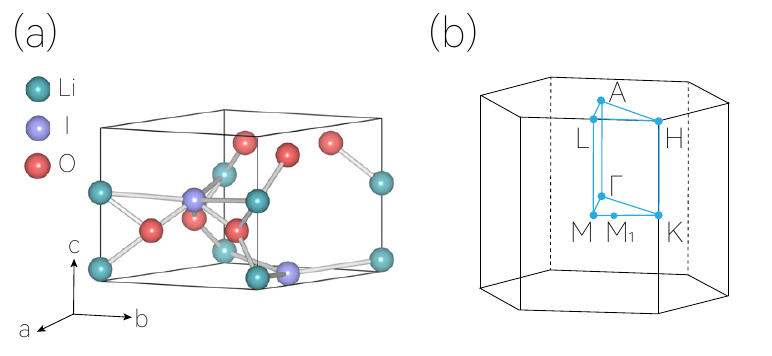}
\caption{(a) Crystal structure of $\alpha$-LiIO$_3$.  (b) Bulk BZ of $\alpha$-LiIO$_3$.
\label{fig2}}
\end{figure}

Consider a 3D system  with $\cal{T}$ and sixfold rotation symmetry along $z$-direction ${\tilde{C}}_{6z,n}=\{C_{6z}|00\frac{n}{6}\}$  ($n=0,1,\cdots, 5$), then the Bloch states on the axis can be chosen as the eigenstates of ${\widetilde{C}}_{6z,n}$,  denoting as
$|(m,n)\rangle$, for which  the  eigenvalue of ${\widetilde{C}}_{6z,n}$  is  $ e^{i2m\pi/6}e^{i k_z n/6}$ ($m=0,1,\cdots, 5$).
While each pair of the bands with different $(m,n)$ can cross and form a WP, the C-3 WP is only formed by the two bands with $|(m,n)\rangle$ and $|(m+3 \  \text{mod}\  6 ,n)\rangle$ \cite{Fang2012Multi-Prl}, as  the ratio of their eigenvalue of ${\widetilde{C}}_{6z,n}$ is $-1$.

\begin{figure*}
\includegraphics[width=18cm]{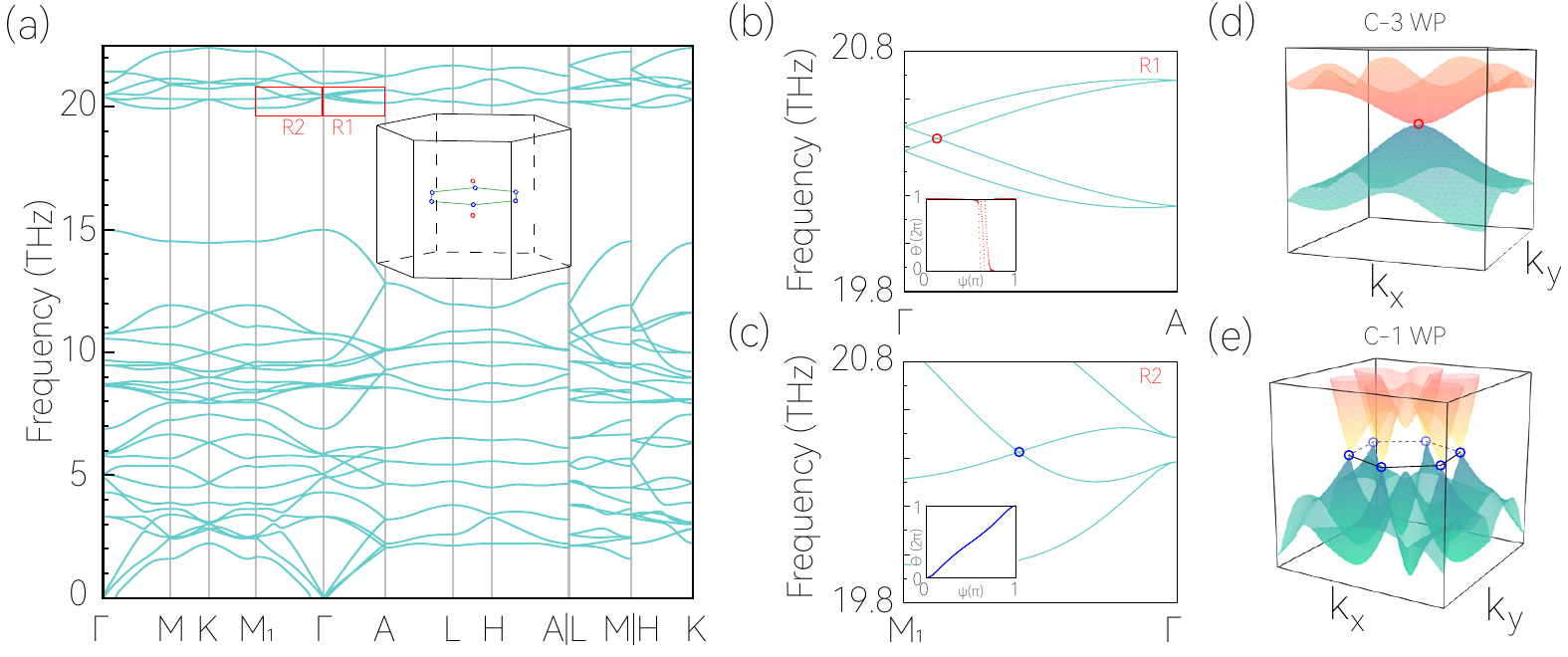}
\caption{(a) Calculated phonon spectrum of  $\alpha$-LiIO$_3$ along high-symmetry paths. The inset in (a) shows the positions of the two C-3 WPs and the six C-1 WPs appearing around 20 THz.
(b) and (c) respectively show the enlarged phonon dispersions of R1 and R2 regions in (a). The hourglass C-3 WP [conventional C-1 WP] can be clearly observed in (b) [(c)]. The inset in (b) [(c)] shows the obtained Wilson loop of the WP in R1  (R2). (d) plots the cubic dispersion around the hourglass C-3 WP in the plane perpendicular to $\Gamma$-A path.
(e) presents the 3D plot of the phonon spectrum at $k_z=0$ plane. The red and blue circles in (a)-(e) show the positions of C-3 and C-1 WPs, respectively.
\label{fig3}}
\end{figure*}

Since ${\cal{T}}^2=1$, at $\Gamma$ $(000)$ and A $(00\pi)$ points, the  state $|(m,n)\rangle$ and its  time-reversal partner ${\cal{T}}|(m,n)\rangle=|(-m,-n)\rangle$ will be  linearly independent when $ e^{i2m\pi/6}e^{i k_z n/6}\neq e^{-i2m\pi/6}e^{-i k_z n/6}$, and must be degenerate at the same energy.
Interestingly, for $n=3$, i.e., ${\tilde{C}}_{6z,3}=\{C_{6z}|00\frac{1}{2}\}$, the two states $|(1,3)\rangle$ and $|(5,3)\rangle$ ($|(2,3)\rangle$ and $|(4,3)\rangle$)  are degenerate at $\Gamma$ point, while  $|(1,3)\rangle$ and $|(2,3)\rangle$ ($|(4,3)\rangle$ and $|(5,3)\rangle$)  are degenerate at $A$ point, leading to an hourglass-type band dispersion, as illustrated in Fig. \ref{fig1}.
Particularly,  the neck crossing-point of the hourglass is formed by the states either $|(2,3)\rangle$ and $|(5,3)\rangle$ [see Fig. \ref{fig1}(a)] or $|(1,3)\rangle$ and $|(4,3)\rangle$ [see Fig. \ref{fig1}(b)], indicating that the neck crossing-point here must be a C-3 WP.
We term such crossing-point as hourglass C-3 WP.
Since only two chiral SGs (SG 173 and SG 182) exhibit $\{C_{6z}|00\frac{1}{2}\}$ (see Table \ref{table I}), the hourglass C-3 WP would be very rare.

\begin{figure*}[htbp]
\includegraphics[width=17cm]{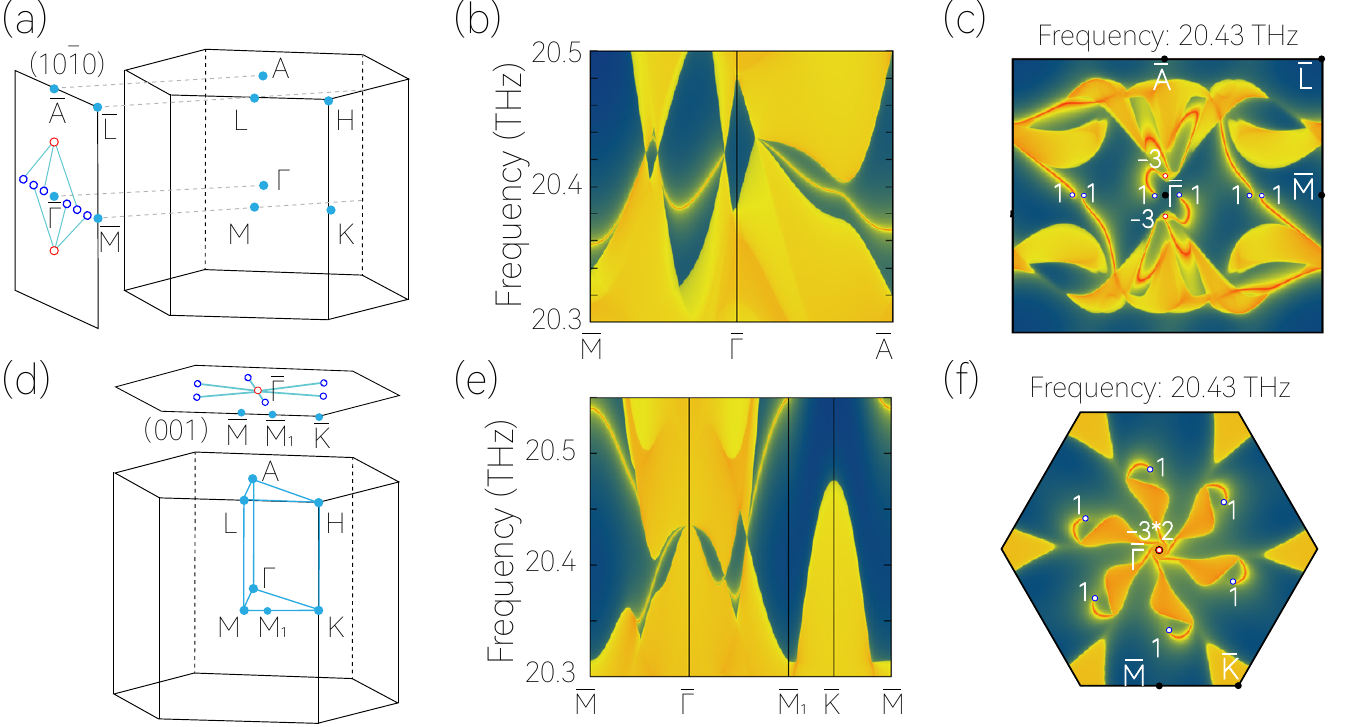}
\caption{(a) and (d) show the $(10\bar{1}0)$  and (001) surface BZs, along with the schematics  of surface modes. The projections of the C-3 and C-1 WPs are marked by red and blue circles, respectively. Projected spectrum on the (b) $(10\bar{1}0)$  and (e) (001) surfaces of $\alpha$-LiIO$_3$. (c) and (f) show the constant energy slices at 20.43 THz for the $(10\bar{1}0)$ and (001) surfaces, respectively.
\label{fig4}}
\end{figure*}

\emph{\textcolor{blue}{Material candidate: $\alpha$-LiIO$_3$.}}
Based on first-principles calculations, we confirm our idea by  predicting   the existing material  $\alpha$-LiIO$_3$ with SG 173 (P6$_3$) as the first material candidate with (hourglass) C-3 Weyl phonons.
$\alpha$-LiIO$_3$ can be prepared by neutralizing stoichiometric quantities of Li$_2$CO$_3$ and HIO$_3$ in distilled water, i.e., Li$_2$CO$_3$+2HIO$_3$ $\rm\rightarrow$ 2LiIO$_3$+H$_2$O+CO$_2$. The solution can be evaporated at 313 K to obtain the $\alpha$-LiIO$_3$ crystallites~\cite{de1966re}.
In our calculation, the structure of $\alpha$-LiIO$_3$ is relaxed, and the optimized crystal structure is shown in Fig. \ref{fig2}(a). The determined lattice constants ($a = b = 5.396$ {\AA} and $c = 5.048$ {\AA}) are in a good agreement with the experimental values~\cite{de1966re}.
The Li, I and O atoms of  $\alpha$-LiIO$_3$ locate at $2a$, $2b$  and $6c$  Wyckoff positions, respectively. More details about the computational methods can be found in SM \cite{SM}.

According to Table \ref{table I}, SG 173 may exhibit an hourglass C-3 WP on $\Gamma$-A path. In Fig. \ref{fig3}(a), we plot the phonon band structure of $\alpha$-LiIO$_3$ along high-symmetry paths in Brillouin zone (BZ), where multiple  hourglass-type bands indeed are observed on $\Gamma$-A path, consistent with above symmetry analysis. Here, we focus on the hourglass-type bands around  $20$ THz, for which the enlarged plot is shown in Fig. \ref{fig3}(b).
The hourglass-type bands are formed by the four phonon branches (No. 25-28). Moreover, one observes that these hourglass-type bands are well separated from other bands [see Fig. \ref{fig3}(b)], which would be beneficial for experimental detections.

We then study the neck crossing-point  of the hourglass in Fig. \ref{fig3}(b), which should be a C-3 WP according to our symmetry analysis. It  locates at $\boldsymbol{K}=(0,0,0.061)$ in the unit of reciprocal lattice vectors. The $k\cdot p$ effective Hamiltonian  of the neck crossing-point expanded up to leading order reads ~\cite{YU2022375}
\begin{eqnarray}
{\cal{H}}_{\text{C-3}}(\boldsymbol{q}) & =  &  {\cal{H}}_0(\boldsymbol{q})+
\begin{bmatrix}0 & \alpha_1 q_+^3+\alpha_2 q_-^3\\
\alpha_1^* q_-^3+\alpha_2^* q_+^3 & 0
\end{bmatrix},\nonumber \\
\label{eq:C-3 WP}
\end{eqnarray}
with
\begin{eqnarray}
{\cal{H}}_0(\boldsymbol{q}) & =  &  \sum_{i=0}^{3}q_z(c_{i,1}+c_{i,2}q_z+c_{i,3}q_z^2+c_{i,4}q^2)\sigma_{i}.
\end{eqnarray}
Here, the energy and momentum $\boldsymbol{q}$ are measured from the neck crossing-point $\boldsymbol{K}$.
$\sigma_{i}$ ($i=0,1,2,3$) is the Pauli matrix, $q_{\pm}=q_x\pm iq_y$, $q=\sqrt{q_x^2+q_y^2+q_z^2}$, and  $\alpha$ and $c$ denote complex and real parameters depending on material details, respectively.
According to Eq. (\ref{eq:C-3 WP}), the band splitting around the neck crossing-point is linear along $q_z$ direction  and cubic  in the $q_x$-$q_y$ plane,  which is confirmed by our calculations [see Fig. \ref{fig3}(d)].
The inset of Fig. \ref{fig3}(b) also plots the Wilson loop of a sphere enclosing the neck crossing-point, and shows its topological charge is  ${\cal{C}}=-3$.
These results undoubtedly demonstrate the  neck crossing-point of the hourglass is a  C-3 WP.

Since the $\alpha$-LiIO$_3$ has ${\cal{T}}$ symmetry,  another C-3 WP with ${\cal{C}}=-3$ will appear at $-\boldsymbol{K}$ point.
Due to the Nielsen-Ninomiya no-go theorem, the net topological charge of the system would  be zero. Therefore, there must exist other WPs formed by the phonon branches No. 26 and No. 27. After a careful scanning, we do find six C-1 WPs at $k_z=0$ plane, connected by ${\tilde{C}}_{6z,3}$, as shown in Fig.~\ref{fig3}(c) and \ref{fig3}(e).
The position for one of  the C-1 WP is $(-0.040,0.310,0)$, and the topological charge for each of the six C-1 WPs  is obtained as ${\cal{C}}=1$ [see Fig. \ref{fig3}(c)].
Hence, the two C-3 WPs and the six C-1 WPs together constitute a  Weyl complex, for which the net topological charge vanishes.
Notice that the Weyl complex here  is different from the triangular Weyl complex proposed by Wang \emph{et al.} in the phonon spectra of $\alpha$-SiO$_2$~\cite{PhysRevLett.124.105303}. There,  the  Weyl complex is composed of one C-2 WP and two C-1 WPs.
Since the six C-1 WPs locate at generic positions of BZ, they do not have ${\cal{T}}$ and  any crystalline symmetry.
Then, the Hamiltonian for each of the  C-1 WP reads
\begin{eqnarray}
{\cal{H}}_{\text{C-1}}(\boldsymbol{q}) & =  &  \sum_{i=0}^{3}(c_{i,1} q_x+c_{i,2}q_y+c_{i,3}q_z)\sigma_{i}, \label{eq:C-1 WP}
\end{eqnarray}
with  $c$'s  the real parameters.
Again,  the energy and momentum in Eq. (\ref{eq:C-1 WP}) are measured from the corresponding C-1 WP.

\emph{\textcolor{blue}{ Triple- and sextuple-helicoid surface arc states.}} In the following, we come to examine the novel  surface states associated with the hourglass C-3 WP in $\alpha$-LiIO$_3$. We first  consider the $(10\bar{1}0)$  surface [see Fig.  \ref{fig4}(a)]. For this surface, the two bulk hourglass C-3 WPs will be projected into the $\bar{\Gamma}$-$\bar{A}$ path in surface BZ , while the six bulk C-1 WPs are projected into the $\bar{\Gamma}$-$\bar{M}$ path, as illustrated  in Fig.  \ref{fig4}(a). Since the two C-3 WPs are projected to  different positions in surface BZ, one can expect that the  three surface arc states of each C-3 WP around the projected point on the boundary would  form a triple helicoid, similar to the other  nodal points with nonzero Chern number \cite{RN1290,PhysRevLett.120.016401,cui2021charge}.
Besides, because the two  C-3 WPs have the same chirality, there in total exist six Fermi arcs on $(10\bar{1}0)$  surface. These Fermi arcs  emerge from  the surface projection of the two C-3 WPs and  end at the surface projections of the six C-1 WPs. Due to the presence of ${\cal{T}}$-symmetry, the surface mode  may  form the pattern in Fig.~\ref{fig4}(a). This analysis on the  surface mode is confirmed by our calculation results in Fig. \ref{fig4}(b) and \ref{fig4}(c), which respectively show the projected spectrum and the iso-frequency surface contour at 20.43 THz of $\alpha$-LiIO$_3$ on the $(10\bar{1}0)$  surface.

In contrast, the surface mode on  the (001) surface shows completely different features. The projected spectrum for $\alpha$-LiIO$_3$ on the (001) surface along the surface paths $\bar{M}$-$\bar{\Gamma}$-$\bar{M_1}$-$\bar{K}$-$\bar{M}$
 is plotted  in Fig. \ref{fig4}(e), and the iso-frequency surface contour at 20.43 THz is shown in Fig. \ref{fig4} (f).  For (001) surface, the two bulk hourglass C-3 WPs are  projected to  a same position, namely,  $\bar{\Gamma}$ point in (001) surface BZ. Thus, there must exist six Fermi arcs  connected to $\bar{\Gamma}$ point, leading to a sextuple helicoid instead of a  triple helicoid, as schematically shown in Fig. \ref{fig4} (d)]. This novel surface mode also is confirmed by our calculation in  Fig. \ref{fig4}(f).

\emph{\textcolor{blue}{Conclusions.}}
In summary, using symmetry analysis, we show that the C-3 WP only appears at sixfold rotation axis and can be further classified  as normal C-3 WP and hourglass C-3 WP.
For the former, it can be realized in 12 out of 230 SGs, while the later only occurs in two SGs.
We then predict  several realistic materials that host hourglass C-3 Weyl phonons. As a representative material candidate, $\alpha$-LiIO$_3$ exhibits one pair of C-3 WPs and six C-1 WPs at certain frequency range in phonon spectra, leading to a novel Weyl complex. Moreover, the triple- and sextuple-helicoid surface arc states can be clearly found in the $(10\bar{1}0)$ surface and (001) surface of $\alpha$-LiIO$_3$, respectively.
Thus, our work uncovers a new type of Weyl phonons, offers a method to search for (hourglass) C-3 Weyl phonons in 230 SGs, proposes realistic materials to realize the ideal hourglass C-3 Weyl phonons and clean sextuple-helicoid phonon surface states. Furthermore, our results are not limited to phonon spectra but can also be applied to other boson systems, such as photon or magnon systems.

\bibliography{refSI}

\end{document}


\title{Hourglass Charge-Three Weyl Phonons}
\author{Xiaotian Wang}
\address{School of Physical Science and Technology, Southwest University, Chongqing 400715, China.}

\author{Feng Zhou}
\address{School of Physical Science and Technology, Southwest University, Chongqing 400715, China.}

\author{Zeying Zhang}
\affiliation{College of Mathematics and Physics, Beijing University of Chemical Technology, Beijing 100029, China}
\affiliation{Research Laboratory for Quantum Materials, Singapore University of
Technology and Design, Singapore 487372, Singapore}

\author{Zhi-Ming Yu}
\affiliation{Centre for Quantum Physics, Key Laboratory of Advanced Optoelectronic Quantum Architecture and Measurement (MOE), School of Physics, Beijing Institute of Technology, Beijing, 100081, China}
\affiliation{Beijing Key Lab of Nanophotonics \& Ultrafine Optoelectronic Systems, School of Physics, Beijing Institute of Technology, Beijing, 100081, China}
\author{Yugui Yao}
\affiliation{Centre for Quantum Physics, Key Laboratory of Advanced Optoelectronic Quantum Architecture and Measurement (MOE), School of Physics, Beijing Institute of Technology, Beijing, 100081, China}
\affiliation{Beijing Key Lab of Nanophotonics \& Ultrafine Optoelectronic Systems, School of Physics, Beijing Institute of Technology, Beijing, 100081, China}
\maketitle


\section{DETAILS OF FIRST-PRINCIPLES CALCULATION}
In this work, we have used the relaxed lattice constants to calculate the phonon spectrums.
Before calculating the phonon spectrums, the atomic positions for the materials are relaxed using the density functional theory within the Vienna $ab\ initio$ simulation package \cite{PhysRevB.54.11169}. The exchange-correlation functional was modeled within the generalized gradient approximation with the Perdew-Burke-Ernzerhof realization \cite{PhysRevLett.77.3865}. The projector augmented-wave method \cite{PhysRevB.50.17953} was selected to deal with the interactions between ions and valence electrons. The  cutoff energy was set to 620 eV for $\alpha$-LiIO$_3$, and 520 eV for NaPO$_3$NH$_3$, Y$_3$CuGeS$_7$, Tl$_3$(PO$_4$), and Zr$_3$O. A $\Gamma$-centered $k$-mesh of $7 \times 7 \times 7$ [$6 \times 6 \times 5$ / $4 \times 4 \times 5$ / $4 \times 4 \times 6$ / $6 \times 6 \times 6$ ] for $\alpha$-LiIO$_3$ [NaPO$_3$NH$_3$/ Y$_3$CuGeS$_7$/ Tl$_3$(PO$_4$)/ Zr$_3$O] was used for Brillouin zone (BZ) sampling. The energy convergence criterion was set to be 10$^{-8}$ eV [10$^{-6}$ eV ] for $\alpha$-LiIO$_3$ [NaPO$_3$NH$_3$, Y$_3$CuGeS$_7$, Tl$_3$(PO$_4$) and Zr$_3$O]. The atomic positions for the materials are totally relaxed until the maximum force on each atom is less than -0.01 eV. Then, we calculate the phonon dispersions of $2 \times 2 \times 2$ $\alpha$-LiIO$_3$, $2 \times 2 \times 2$ NaPO$_3$NH$_3$, $2 \times 2 \times 2$ Y$_3$CuGeS$_7$, $2 \times 2 \times 2$ Tl$_3$(PO$_4$), and $2 \times 2 \times 2$ Zr$_3$O supercells by density functional perturbation theory using the PHONOPY package  \cite{TOGO20151}. The phonon surface states were calculated by constructing a Wannier tight-binding Hamiltonian of phonons \cite{WU2018405}.

\begin{figure}
\includegraphics[width=14cm]{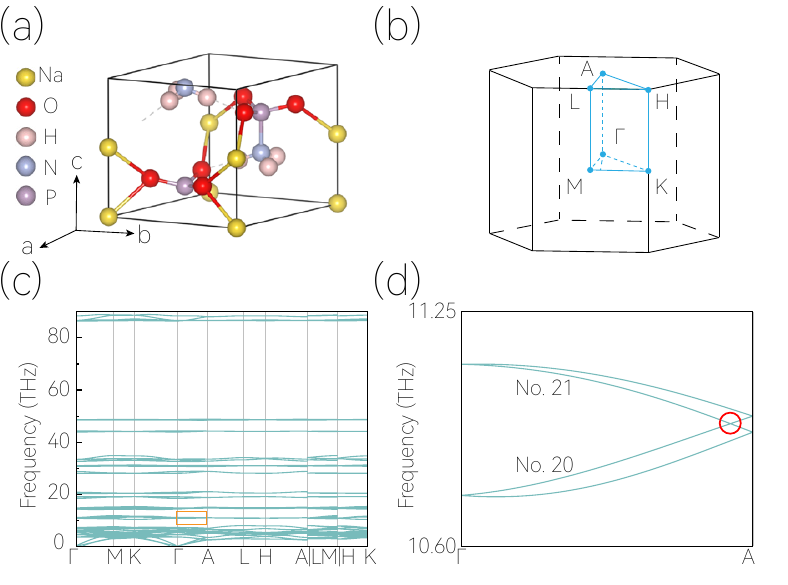}
\caption{ (a) crystal structure of NaPO$_3$NH$_3$ with SG 173 and obtained lattice constants $a = b = 5.797$ {\AA} and $c = 6.179$ {\AA}. (b) 3D BZ. (c) calculated phonon spectrum of NaPO$_3$NH$_3$ along high-symmetry paths. (d) shows the enlarged phonon dispersions in the region marked by orange box in (c),  where the hourglass C-3 Weyl phonon (marked by red circle) can be clearly observed on $\Gamma$-A path.
\label{figS1}}
\end{figure}

\begin{figure}
\includegraphics[width=14cm]{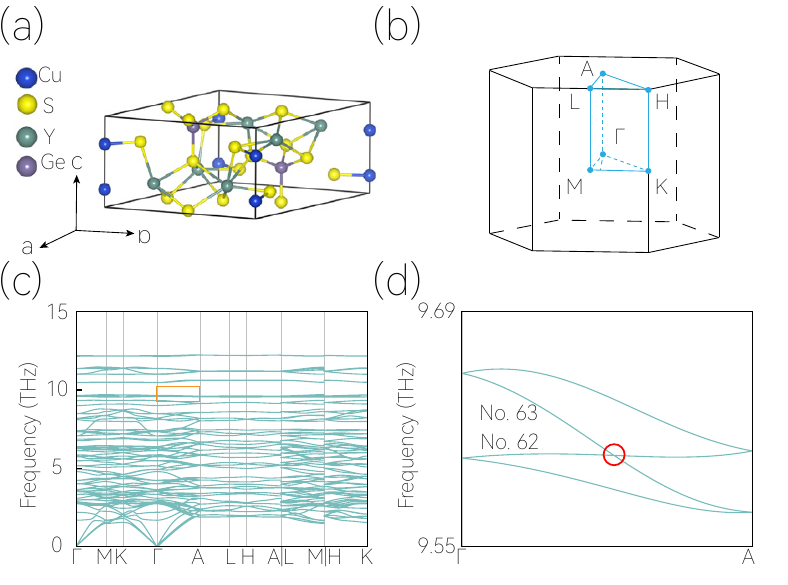}
\caption{ (a) crystal structure of Y$_3$CuGeS$_7$ with SG 173 and obtained lattice constants $a = b = 9.877$ {\AA} and $c = 5.839$ {\AA}. (b) 3D BZ. (c) calculated phonon spectrum of Y$_3$CuGeS$_7$ along high-symmetry paths. (d) shows the enlarged phonon dispersions in the region marked by orange box in (c),  where the hourglass C-3 Weyl phonon (marked by red circle) can be clearly observed on $\Gamma$-A path.
\label{figS2}}
\end{figure}

\begin{figure}
\includegraphics[width=13cm]{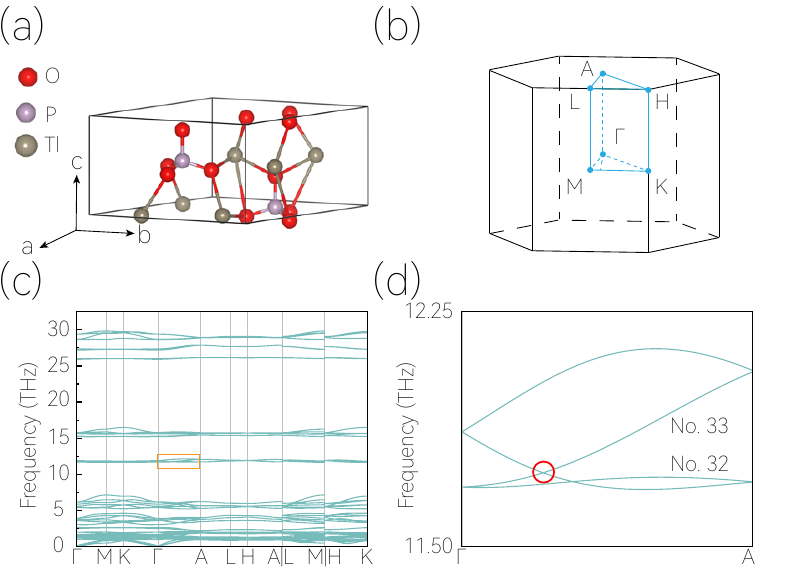}
\caption{ (a) crystal structure of Tl$_3$(PO$_4$) with SG 173 and obtained lattice constants $a = b = 8.366$ {\AA} and $c = 5.086$ {\AA}. (b) 3D BZ. (c) calculated phonon spectrum of Tl$_3$(PO$_4$) along high-symmetry paths. (d) shows the enlarged phonon dispersions in the region marked by orange box in (c),  where the hourglass C-3 Weyl phonon (marked by red circle) can be clearly observed on $\Gamma$-A path.
\label{figS3}}
\end{figure}

\begin{figure}
\includegraphics[width=13cm]{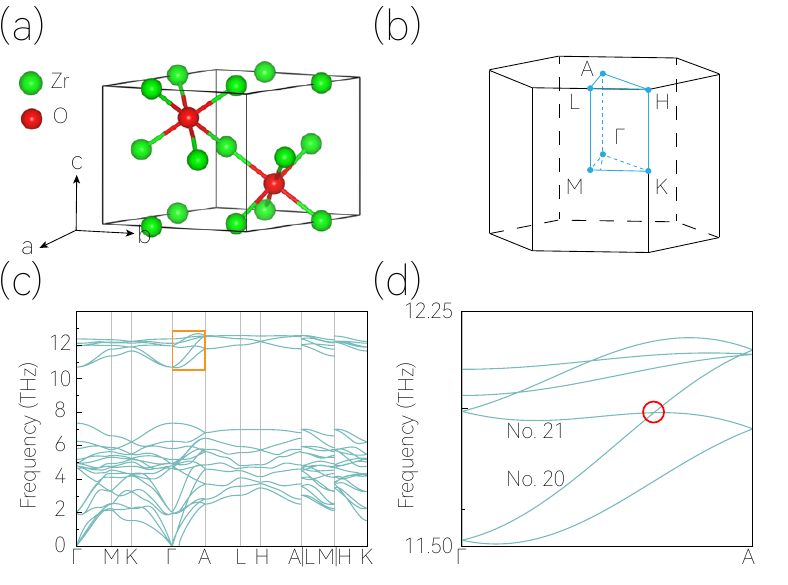}
\caption{ (a) crystal structure of Zr$_3$O with SG 182 and obtained lattice constants $a = b = 5.651$ {\AA} and $c = 5.214$ {\AA}. (b) 3D BZ. (c) calculated phonon spectrum of Zr$_3$O along high-symmetry paths. (d) shows the enlarged phonon dispersions in the region marked by orange box in (c),  where the hourglass C-3 Weyl phonon (marked by red circle) can be clearly observed on $\Gamma$-A path.
\label{figS4}}
\end{figure}

\bibliography{refSII}